\documentclass[twocolumn,pre,amsmath,amssymb]{revtex4}

\usepackage{graphicx}
\usepackage{bm}

\begin{document}

\title{Power law relaxation and glassy dynamics in Lebwohl-Lasher model near isotropic-nematic phase transition}

\author{Suman Chakrabarty}
\author{Dwaipayan Chakrabarti}
\author{Biman Bagchi}
\email{bbagchi@sscu.iisc.ernet.in}
\homepage{http://liquid.sscu.iisc.ernet.in/}
\affiliation{Solid State and Structural Chemistry Unit,\\
Indian  Institute of Science, Bangalore 560012, India.}

\date{\today}

\begin{abstract}

Orientational dynamics in a liquid crystalline system near the isotropic-nematic (I-N) phase 
transition is studied using Molecular Dynamics simulations of the well-known 
Lebwohl-Lasher (LL) model. As the I-N transition temperature is approached from the isotropic side,
we find that the decay of the orientational time correlation functions (OTCF) slows down 
noticeably, giving rise to a power law decay at intermediate timescales. 
The angular velocity time correlation function also exhibits a rather pronounced 
power law decay near the I-N boundary. In the mean squared angular displacement 
at comparable timescales, we observe 
the emergence of a \emph{subdiffusive regime} which is followed by a 
\emph{superdiffusive regime} before the onset of the long-time diffusive behavior. We observe 
signature of dynamical heterogeneity through \emph{pronounced non-Gaussian behavior 
in orientational motion} particularly at lower temperatures. This behavior closely 
resembles what is usually observed in supercooled liquids.
We obtain the free energy as a function of orientational order parameter by the use of 
transition matrix Monte Carlo method. The free energy surface is flat for the 
system considered here and the barrier between isotropic and nematic phases 
is vanishingly small for this weakly first-order phase transition, hence 
allowing large scale, collective and correlated orientational density fluctuations. 
This might be responsible for the observed power law decay of the OTCFs.

\end{abstract}



\maketitle

\section{\label{sec:intro}Introduction}

Liquid crystalline systems often exhibit interesting dynamics apart from the 
rich phase behavior. Surprisingly, dynamics of such systems have 
traditionally been probed only at rather long timescales (nanoseconds to 
milliseconds)~\cite{deglc, lc:chansekh, lc:oswald:pieranski}. 
Recently, Fayer and coworkers have investigated the dynamics in the isotropic 
phase of thermotropic liquid crystals over a wide range of timescales using optical 
Kerr effect (OKE) measurements~\cite{fayer:bagchi:jcp:2002:360, 
fayer:bagchi:jcp:2002:6339, cang:fayer:cpl:2002, cang:fayer:jcp:2003}. At short to 
intermediate timescales, they have observed a pronounced power law 
decay of the time-dependent OKE signal near the I-N phase transition. At 
the intermediate times (several nanoseconds) the decay becomes 
even slower, almost appearing as a plateau on a log-log scale~\cite{fayer:bagchi:jcp:2002:360}. 
The exponential decay predicted by Landau-de Gennes theory is observed 
only at the longest timescale ($>10$ ns).

Subsequent molecular dynamics simulations of a calamitic system (comprising 
of rod-like molecules) with the Gay-Berne pair potential were found to 
reproduce the power law decay of orientational time correlation functions 
(OTCF)~\cite{jose:bagchi:jcp:2004:11256}. From a very recent computational study, 
which deals with a calamitic 
system, a discotic system (comprising of disc-like molecules) and to a limited 
extent the Lebwohl-Lasher lattice model, it appears that this power law decay 
at short-to-intermediate times might be a rather general phenomenon in thermotropic 
liquid crystals~\cite{liqcrys:prl:2005}. This observation has gained support from 
a recent finding of power law decay at relatively short times in an idealized 
calamitic liquid crystal model with length-to-width ratio 
5-6~\cite{bertolini:tani:calamitic:model}. It has been observed that many aspects 
of the orientational relaxation behavior outlined above bear close resemblance 
to what is observed in supercooled molecular liquids near the glass transition 
temperature~\cite{cang:fayer:jcp:2003, jose:chakrabarti:bagchi:pre:2005:030701, liqcrys:prl:2005}. 
In particular, the description of orientational relaxation with a power law decay 
at short times and exponential decay at long times is strikingly similar~\cite{cang:fayer:jcp:2003}. 
However, the origin of such a rich dynamical behavior may be quite different in the 
two cases~\cite{liqcrys:prl:2005}. It would be especially interesting to explore 
the free energy surface with respect to orientational density fluctuations 
in search of the origin of the slow dynamics near the I-N transition.

Comprehensive understanding of this rather exotic dynamics of thermotropic 
liquid crystals spread over almost five decades of time is a challenging task, 
particularly because of the anisotropic nature of the interaction, which makes the 
theoretical analysis difficult. Computational approaches have become very much useful 
in this regard as we gain control over the microscopic interactions and try to 
understand their manifestation into the macroscopic behavior. In this spirit, 
this communication attempts to continue the investigation of the Lebwohl-Lasher 
model.

Lebwohl-Lasher (LL) model~\cite{leblash} is essentially the lattice version of the Maier-Saupe 
model. In this model, the molecules being fixed on a simple cubic 
lattice lose their translational degrees of freedom and can only rotate. 
The total interaction energy is given as follows:
\begin{equation}
\label{leblash:potential}
  U = - \frac{1}{2}\sum_{i, j}\epsilon_{ij} (\frac{3}{2} \cos^2 \theta_{ij} - \frac{1}{2})
\end{equation}
where $\epsilon_{ij}$ is the coupling parameter, signifying the strength of
interaction between the rotors. It has a fixed value $\epsilon$ for the nearest 
neighbors and $0$ otherwise.

LL model has been rather popular since its inception in computer simulation
studies of liquid crystals. Its merit lies in the inherent simplicity with which it
clearly establishes some of the very essential features of orientational ordering in
liquid crystalline systems~\cite{leblash:md:zannoni, leblash:langdyn, greef:lee:pre:49:3225, 
ll_freenrg:prl:1992, pas:zan}. Simple form of the anisotropic interaction and the absence of
translational degrees of freedom makes the system particularly easy to study.
Most of the earlier studies involving LL model used Monte Carlo (MC) methods, which  
can not predict the real dynamics of the system. In view of this, here we have undertaken 
molecular dynamics (MD) simulations to investigate the dynamics in the LL model. 

Our MD simulations of the LL model show that as the transition temperature is approached from 
the isotropic side, the decay of the orientational time correlation functions slows 
down noticeably, giving rise to power law decay at intermediate timescales. Another 
interesting result is the emergence of \emph{subdiffusive orientational motion} 
essentially in the same temporal window where the power law decay sets in. 
In addition, the sub-diffusive motion is followed by a \emph{superdiffusive regime} 
before the long-time diffusive behavior sets in. Moreover, we have used the recently 
developed \emph{transition matrix Monte Carlo} (TMMC) method to obtain the free energy profile
for the system as a function of orientational order parameter. We find the 
barrier to be vanishingly small for this
weakly first-order transition allowing large fluctuations in the orientational order.

The organization of the rest of the paper is as follows. In the next section, 
we discuss the details of simulations. In section~\ref{results}, we present the results on
orientational relaxation. Section~\ref{tmmc} deals with the TMMC method and the results 
it yields on the free energy calculation. Section~\ref{conclusion} concludes 
with summary and a few pertinent comments.

\section{Simulation details}

The system under study consists of a $(10 \times 10 \times 10)$ simple cubic lattice
with one rotor fixed on every lattice point. Only the nearest neighbors 
interact through the orientation dependent 
potential given by Eq.~(\ref{leblash:potential}). Throughout we have used
dimensionless temperature $T^{\ast}$ and time $t^{\ast}$ (for the MD studies)
defined by~\cite{leblash:md:zannoni}:

\begin{eqnarray}
\label{dimension}
T^{\ast} & = & k_{B}T/\epsilon \nonumber \\
t^{\ast} & = & t(\epsilon/I_{\perp})^{1/2}
\end{eqnarray}

Here both the parameters $\epsilon$ and $I_{\perp}$ (moment of inertia
with respect to the axis perpendicular to the molecular axis) have been taken
to be unity.

We have performed Molecular Dynamics (MD) simulations in microcanonical (NVE) 
ensemble using velocity Verlet algorithm. We have used time step of $\delta t^{\ast} = 0.002$ 
in reduced unit and have obtained energy conservation upto fifth place 
of decimal throughout the simulation. We have scaled the velocities at every $100$ steps 
for initial $10^{5}$ steps to equilibrate the 
system at a particular desired temperature and stored the trajectory for analysis after 
allowing the system to evolve without scaling for another $10^{5}$ steps. 
The standard deviation in temperature 
during data acquisition was of the order of $0.01$ for all temperatures. 
Periodic Boundary Conditions (PBC) have been applied to remove the surface effects. Earlier MC studies 
showed pronounced system size dependence for this model. 
The same applies for the MD simulation also. The size effect is particularly important 
near the transition temperature as the correlation length tend to diverge. The transition 
becomes sharper with increasing system size. The transition temperature ($T_{IN}$) 
and the free energy barrier follow certain finite size scaling laws~\cite{ll_freenrg:prl:1992}. 
We have found the dynamical features reported here to be qualitatively similar 
for different system sizes.

\section{Results}
\label{results}

\begin{figure}
\centering
\includegraphics[width=0.47\textwidth]{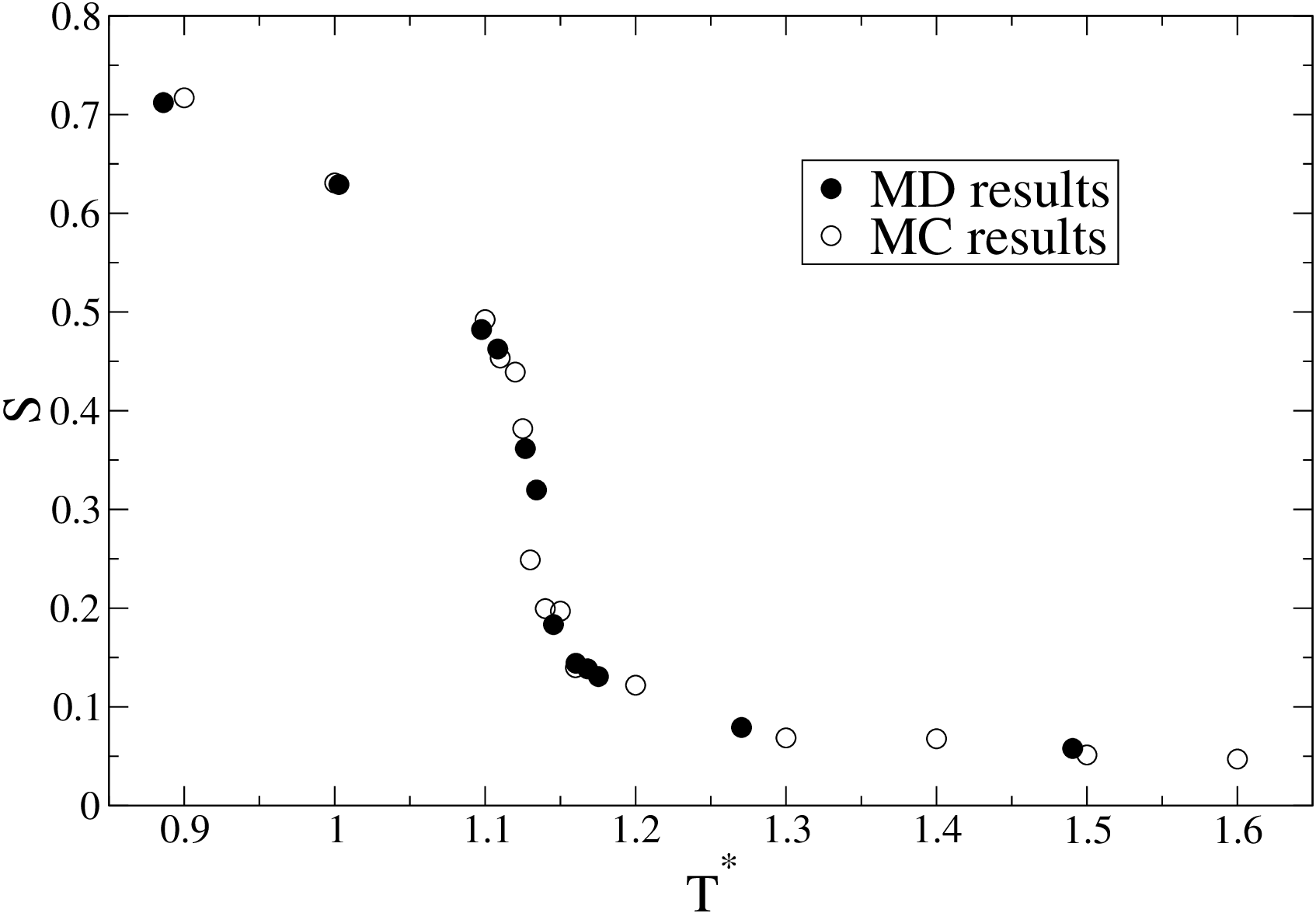}
\caption{\label{fig:ordpar} The average orientational order parameter indicating 
the I-N transition at $T^{*}$ = 1.14. Empty circles indicate the MC results and 
filled circles indicate the MD results.}
\end{figure}

The phase behavior for LL model has been well explored using Monte Carlo (MC) methods 
for pretty large systems. But since there are relatively less number of Molecular 
Dynamics (MD) studies, we have computed the average order parameter at various 
temperatures using MD trajectories. The value of orientational order parameter is determined 
by diagonalization of the ordering matrix $\bm{Q}$~\cite{deglc, pas:zan}:

\begin{equation}
\label{3d:tensor}
  \bm{Q}_{\alpha \beta}=\frac{1}{2N} \sum_{i=1}^{N}
	[3e_{i\alpha}e_{i\beta} - \delta_{\alpha \beta}],
\end{equation}

where $e_{i\alpha}$ is the $\alpha$-th Cartesian coordinate of the unit
vector ($\bm{e}_{i}$) specifying the orientation of the $i$-th molecule.
The largest eigenvalue and corresponding eigenvector give the 
orientational order parameter and the director respectively for a particular 
configuration. The thermodynamic order parameter ($S$) is obtained by averaging over the simulation 
trajectory. As shown in Fig.~\ref{fig:ordpar} the data obtained 
from the MD simulation overlaps well with the result obtained from our MC simulation.

Note that MD simulation gives us the value of the transition temperature to be
$\simeq 1.14$. This differs from the more accurate value ($\simeq 1.1232$) 
obtained from MC simulations~\cite{leblash:mc:zannoni:molphys, ll_freenrg:prl:1992} at
the second place of decimal. This is acceptable because of the fluctuation in 
temperature present in microcanonical (NVE) MD simulation and the standard deviation of the order
of $0.01$ at all temperatures. The fluctuation in temperature is 
particularly high near $T_{IN}$. We have used the value
obtained from our MD simulation to explain our other observations.

\begin{figure}
\centering
\includegraphics[width=0.47\textwidth]{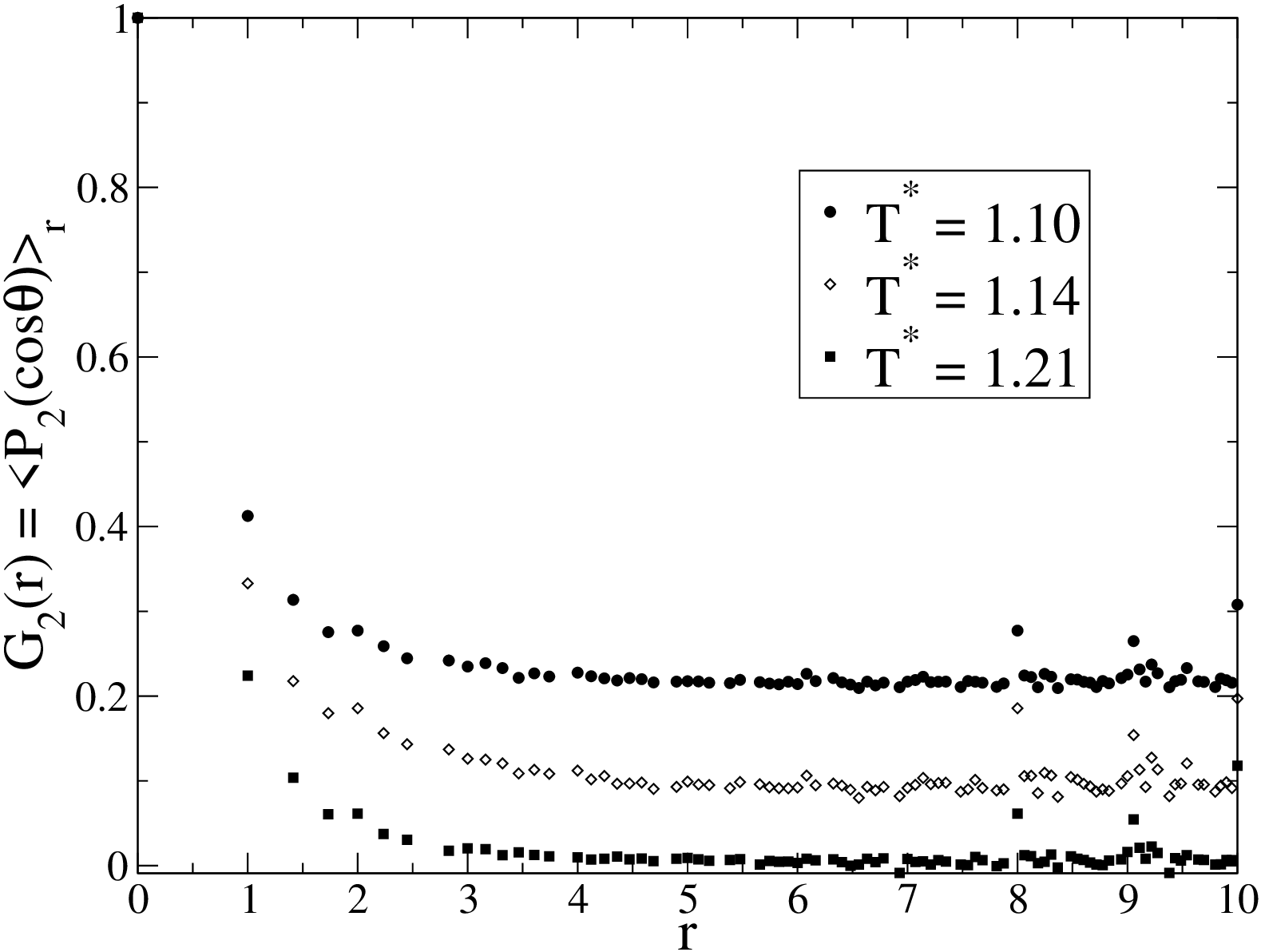}
\caption{\label{fig:G2r} The nature of spatial decay of the two particle orientational 
correlation function $G_{2}(r)$ clearly demonstrates the existence 
of long range order in nematic phase. The correlation function decays to very small value as 
isotropic phase is reached.}
\end{figure}

Often the second rank order parameter is not enough to decide if true long range 
order exists in the system. For that purpose, orientational correlation functions ($G_{l}(r)$) are 
particularly useful. The set of correlations can be defined as expansion 
coefficients of the rotationally invariant pair distribution~\cite{pas:zan}:

\begin{equation}
  G(r, \beta_{12}) = G_{0}^{00}(r)\sum_{l}\frac{2l+1}{64\pi^{2}}
	G_{l}(r)P_{l}(\cos \theta_{ij}),
\end{equation}

where $G_{0}^{00}(r)$ is the particle center distribution, $\theta_{ij}$ is 
the angle between rotors $i$ and $j$, and $P_{l}$ is the Legendre polynomial of rank $l$. 
For a simple cubic lattice,

\begin{equation}
   G_{0}^{00}(r) = \frac{1}{4\pi \rho r^{2}}\sum_{k}z_{k}\delta(r-r_{k}),
\end{equation}

where $\rho$ is the density and $z_{k}$ is the number of neighbors at $r_{k}$.
Hence, $G_{l}(r) = \langle P_{l}(\cos \theta_{ij})\rangle_{r}$
gives orientational correlation between two rotors separated
by distance $r$. Figure~\ref{fig:G2r} demonstrates that if true long range order
is present, $G_{2}(r)$ decays to a plateau with value $\langle P_{2}\rangle$~\cite{pas:zan}.

\subsection{Power law decay in OTCFs}

Single particle OTCF gives a temporal measure
of the loss of the memory of a single particle of its own orientation in the environment created
by the surrounding molecules. The single particle OTCF of rank $l$ is defined as:
\begin{equation}
  C_{l}^{s}(t) = \frac{\langle \sum_{i}P_{l}(\bm{e}_{i}(0)\cdot \bm{e}_{i}(t))\rangle}
	{\langle \sum_{i}P_{l}(\bm{e}_{i}(0)\cdot \bm{e}_{i}(0))\rangle}
\end{equation}
where $\bm{e}_{i}(t)$ is the unit vector denoting the orientation of
$i$-th molecule at time $t$. Since LL model has up-down symmetry, 
$C_{2}^{s}(t)$ would be physically meaningful.

\begin{figure}
\centering
\includegraphics[width=0.47\textwidth]{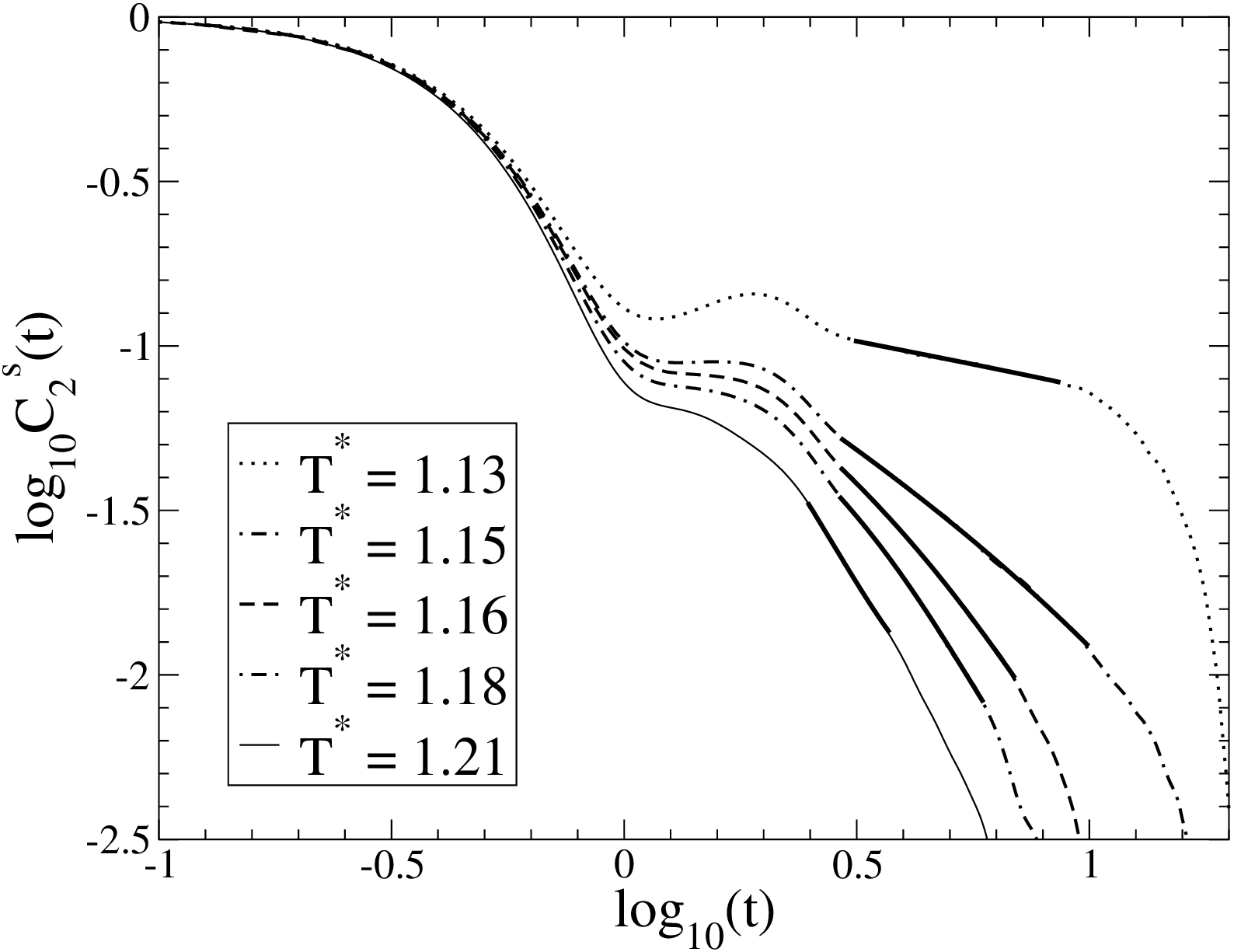}
\caption{\label{fig:singcorr} Demonstration of power law regimes in the log-log plot
of single particle orientational correlation function $C_{2}^{s}(t)$: Fitted power law 
regimes are shown in thick lines. The decay is almost exponential for the highest two 
temperatures, i.e. the power law component becomes very small.}
\end{figure}

In Fig.~\ref{fig:singcorr} the log-log plot of $C_{2}^{s}(t)$ versus time is 
shown at different temperatures approaching $T_{IN}$. 
An interesting step-like decay is clearly evident. The power law behavior emerges
just above the transition temperature, i.e. in the isotropic phase close to the I-N 
transition and continues in the 
nematic phase. It has been rather difficult to fit the entire decay by a single function due to the 
oscillatory regime in the nematic phase and rather dominant plateau appearing 
just before the power law regime. Hence, the OTCF has been fitted only beyond 
the initial plateau using the following functional form.
\begin{equation}
  C_{2}^{s}(t) = A + \exp (-t/\tau) (t/\tau_{p})^{-a}
\end{equation}
Here, the parameter $\tau$ gives the timescale corresponding to 
the long time  exponential decay, the parameter $\tau_{p}$ signifies the timescale 
of the power law decay and the parameter $a$ gives the power law exponent. The values
for $\tau$, $\tau_{p}$ and $a$ are shown in the following table. Note that the 
nature of the decay of the single-particle OTCF is strikingly similar to the 
experimental results obtained by OKE measurements.

\begin{center}
\begin{tabular}{|c|c|c|c|} \hline
$T^{\ast}$	& $\tau$	& $\tau_{p}$	& $a$			\\ \hline \hline
1.13		& 66.17		& 0.002		& 0.33			\\ \hline \hline
1.15		& 12.86		& 0.1		& 0.82			\\ \hline
1.16		& 3.67		& 0.07		& 0.64			\\ \hline
1.18		& 1.69		& $\simeq$ 0.0	& 0.001			\\ \hline
1.21		& 0.84		& $\simeq$ 0.0	& $\simeq$ 0.0		\\ \hline
\end{tabular}
\end{center}

We have an interesting observation that the timescale of the exponential 
decay at the longer times becomes progressively  larger with decreasing temperature, whereas the values 
of $\tau_{p}$ clearly indicate that power law decay exists only near I-N phase boundary and  this
becomes transient as one moves away from the transition region. 
The values of the exponent $a$ follows a similar 
trend. The OTCFs for the highest two temperatures are almost purely exponential beyond the plateau.
We interpret the above result in the following fashion. 
As the temperature approaches $T_{IN}$ from the isotropic side, 
the single particle orientational relaxation
becomes slower due to the strong coupling of rotational motion with
the surrounding molecules, which result in an orientational caging
effect of the rotors. The formation of orientational cage
or \emph{pseudo-nematic domains} (within the isotropic phase) surrounding a molecule, 
gives rise to the retention of the memory of the previous orientation 
for a longer time, resulting in the slow
relaxation process. Thus, the collective relaxation becomes important near
the transition point.

Mode coupling theory (MCT) can be used to obtain a semi-quantitative theoretical 
description and also physical insight of the slow dynamics near phase 
transitions~\cite{ma:mazenko:1975, hohenberg:halperin, sarika:bagchi:acp}. 
MCT starts with splitting the frequency $(z)$ dependent memory function, 
here the rotational friction, into two parts:
\begin{equation}
  \Gamma_{R}(z) = \Gamma_{RB}(z) + \Gamma_{R\rho}(z)
\end{equation}
where $\Gamma_{RB}(z)$ is the short range or binary part which decays on a short 
timescale and usually originates from collisions between the molecules. Whereas, 
$\Gamma_{R\rho}(z)$ is the collective part, deriving contribution from collective 
correlations.

In the present lattice model there is no collisional contribution
to the rotational friction. Therefore, $\Gamma_{RB}(z)$ can be set to zero and this is in fact 
the reason for the large contribution of the initial inertial decay. Friction is small at short times. 
However, $\Gamma_{R\rho}(z)$ exhibits singular features at small frequencies due to 
the emergence of long range orientational correlations near the I-N phase boundary. 
It was shown elsewhere that at low frequency, the frequency dependent friction develops 
a rapid growth which can be represented as~\cite{fayer:bagchi:jcp:2002:360}:
\begin{equation}
\label{memory:powerlaw}
  \Gamma(z) \simeq A z^{-\alpha}
\end{equation}
Mean field treatment gives $\alpha = 0.5$. In general, invoking the rank ($l$) dependence 
of the memory function, the single particle OTCF can be expressed 
as~\cite{ravi:bagchi:1995, chandra:bagchi:acp, hubbard:wolynes:1978, bagchi:jml:1998}:
\begin{equation}
  C_{l}^{s}(z)=\left[z+\frac{l(l+1)k_{B}T}{I(z+\Gamma_{l}(z))}\right]^{-1}
\end{equation}
As discussed by Gottke \textit{et al}, this expression can be Laplace inverted to 
obtain a power law decay in $C_{2}^{s}(t)$ over a range of 
timescales~\cite{fayer:bagchi:jcp:2002:360}. Note that Eq.~(\ref{memory:powerlaw}) 
is valid neither at large nor at very small $z$, rather at intermediate times. 

To study the collective relaxation, we have calculated the collective OTCF defined as:
\begin{equation}
  C_{l}^{c}(t) = \frac{\langle \sum_{i} \sum_{j} P_{l}(\bm{e}_{i}(0)\cdot \bm{e}_{j}(t))\rangle}
	{\langle \sum_{i} \sum_{j} P_{l}(\bm{e}_{i}(0)\cdot \bm{e}_{j}(0))\rangle}
\end{equation}
Evidently, the calculation becomes computationally quite expensive even for 
a 1000 particle system.

\begin{figure}
\centering
\includegraphics[width=0.47\textwidth]{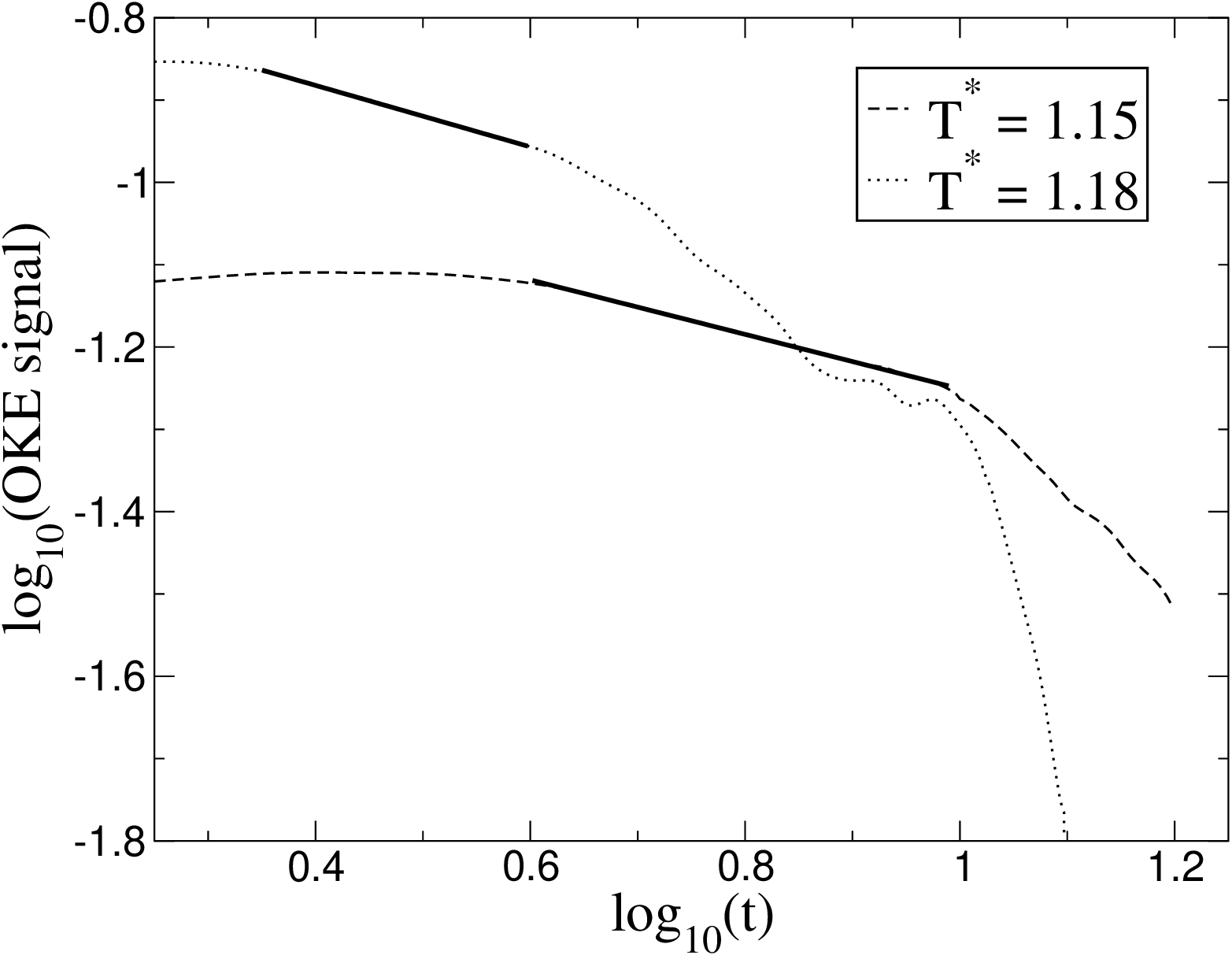}
\caption{\label{fig:collcorr} Power law is exhibited in the derivative of collective 
OTCF, which is essentially the OKE signal. Fitted power law regimes are shown by thick lines.}
\end{figure}

Note that the time derivative of the collective OTCF is directly related to the time 
derivative of polarizability-polarizability correlation function and hence, to the 
OKE signal~\cite{fayer:bagchi:jcp:2002:360}.
In Fig.~\ref{fig:collcorr} we show the prominent power law regimes in the log-log 
plot of derivative of $C_{2}^{c}(t)$ at two temperatures just 
above $T_{IN}$. In the case of the collective OTCF, the power law 
exponents vary only little with temperature. The value of the 
exponents are 0.33 for $T^{\ast} = 1.15$ and 0.37 for $T^{\ast} = 1.18$.

Experimental studies on liquid crystalline systems with rod-like molecules 
find the value of the power law exponent in the range of 0.6-0.7~\cite{cang:fayer:jcp:2003}. 
As observed before, the values of the exponent $a$ are not universal.

\subsection{Power law decay in angular velocity autocorrelation function}

\begin{figure}
\centering
\includegraphics[width=0.47\textwidth]{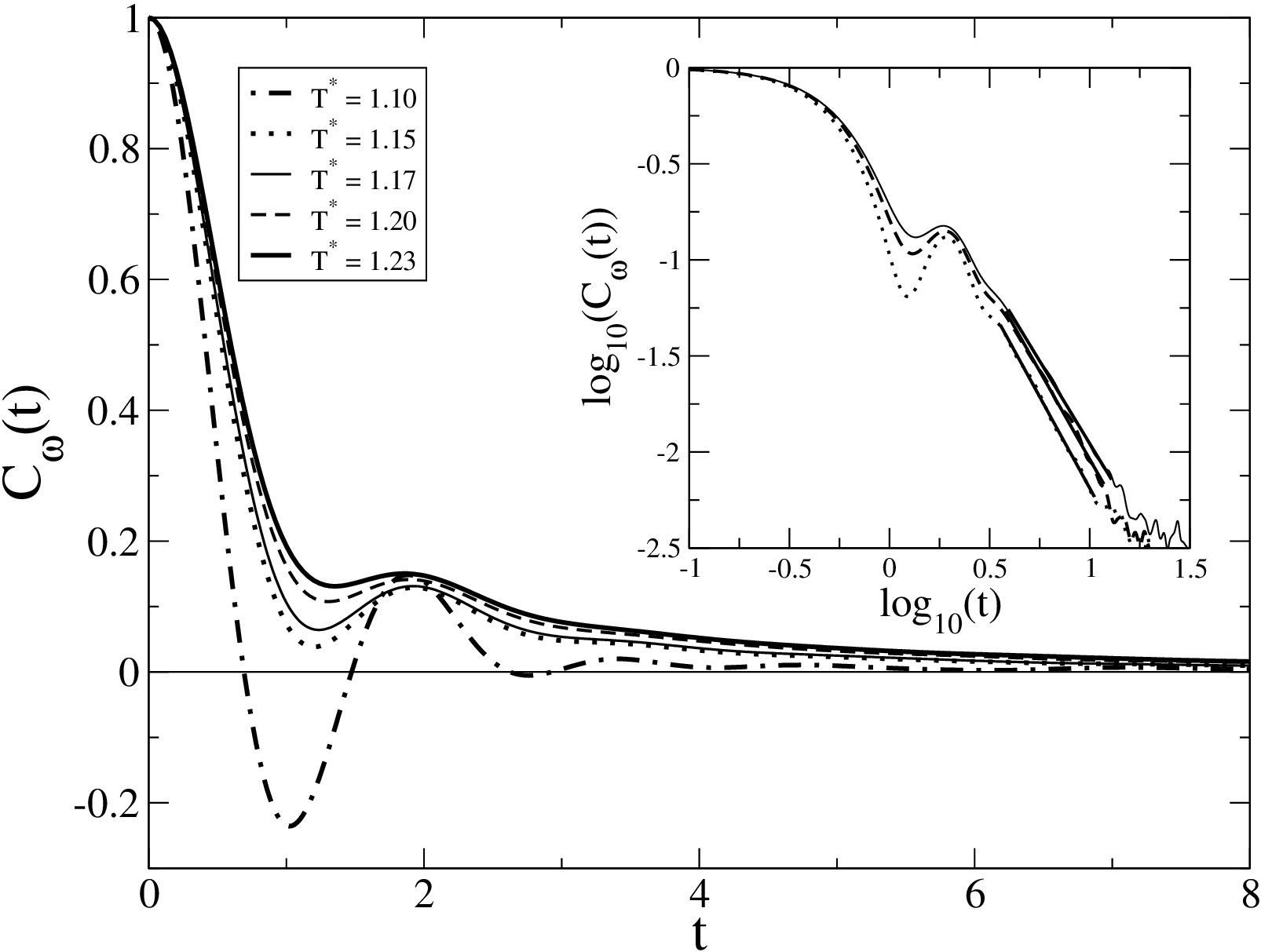}
\caption{\label{fig:angvelcorr} Power law is exhibited in the angular velocity 
autocorrelation function at five temperatures across I-N phase transition. 
Inset figure shows the log-log plot for three temperatures and the thick lines 
indicate the linear fits obtained in power law regimes.}
\end{figure}

We have calculated the angular velocity autocorrelation function $(C_{\omega}(t))$ 
defined as:
\begin{equation}
  C_{\omega}(t) = \frac{\langle \omega(0) \cdot \omega(t) \rangle}{\langle \omega(0) \cdot \omega(0) \rangle}
\end{equation}
where the angular velocity $\omega(t)$ is perpendicular to the axis of the rotor. 
In Fig.~\ref{fig:angvelcorr} we have shown the behavior of $(C_{\omega}(t))$ at 
different temperatures near $T_{IN}$. We find that $(C_{\omega}(t))$ has an 
oscillatory feature at short timescale due to the 
underdamped nature of the system. But we observe a pronounced power law decay 
with the value of the exponent being in the range of 1.7-1.8 at longer timescales. 
The origin of this power law can be attributed to the collective and correlated 
orientational density fluctuations.

As in the case of orientational correlation function $C_{2}^{s}(t)$, a semi-quantitative 
understanding of $C_{\omega}(t)$ can be obtained from MCT. Since the rotational friction 
is given by Eq.~(\ref{memory:powerlaw}), we have the following approximate expression for 
$C_{\omega}(z)$:
\begin{equation}
  C_{\omega}(z) = \frac{k_{B}T}{I(z+Az^{-\alpha})}
\end{equation}
This expression also gives rise to the power law decay of $C_{\omega}(t)$ at long times. 
Note that decay of $C_{\omega}(t)$ occurs at shorter times than that of $C_{2}^{s}(t)$. 
This is the reason for lower amplitude of power law decay in $C_{\omega}(t)$.

\subsection{Rotational diffusion}

To probe the origin of the slow dynamics near the I-N transition, we have studied the
rotational diffusion of the rotors on the lattice. The rotational displacement
has been computed by integrating the angular velocity to have an unbound representation
~\cite{schilling:rotation:supercool,jose:chakrabarti:bagchi:pre:2005:030701}.

\begin{equation}
\vec{\phi}_{n}(t) - \vec{\phi}_{n}(0) = \int_{0}^{t}\,dt' \vec{\omega}_{n}(t')
\end{equation}

\begin{figure}
\centering
\includegraphics[width=0.47\textwidth]{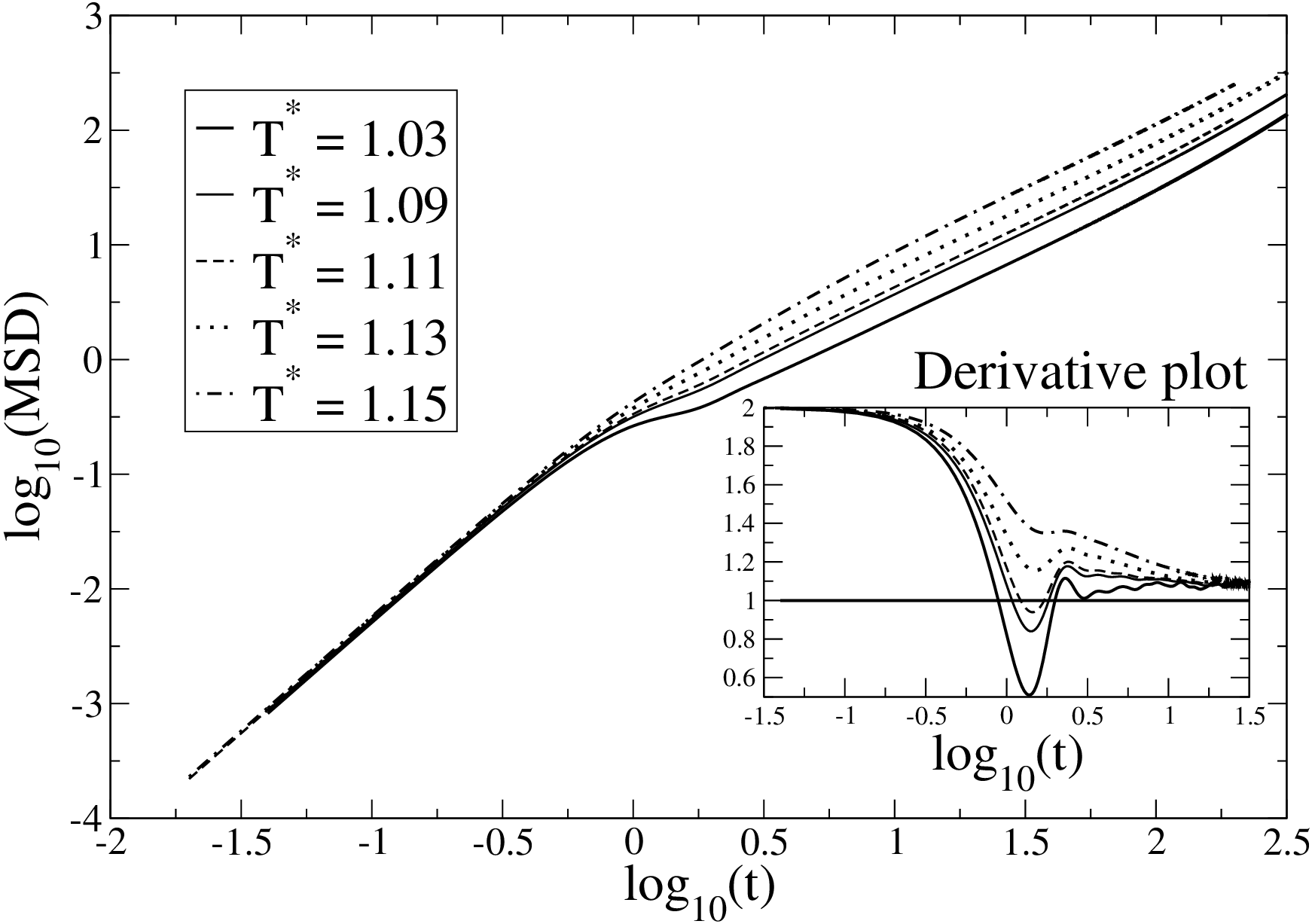}
\caption{\label{fig:msd} The log-log plot of the rotational mean square displacement 
shows the gradual onset of subdiffusive regime followed by a superdiffusive 
jump as temperature is lowered across I-N phase boundary. The derivative of the 
same plot has been shown in the inset figure to clearly demonstrate the striking 
behavior. At very long time the derivative should converge to 1 corresponding to the 
diffusive limit.}
\end{figure}

In Fig.~\ref{fig:msd} the log-log plot of the mean squared angular displacement 
versus time has been shown. Interestingly, we observe the emergence of a 
subdiffussive regime at a timescale comparable
to the plateau observed in single particle OTCF. \emph{This behavior becomes
apparent even in the isotropic phase} and has been illustrated by the 
dip in the derivative of the log-log plot (inset of Fig.~\ref{fig:msd}). 
The derivative plot shoots up suddenly after the dip and the motion remains 
superdiffussive. The system takes considerably long time to attain the 
diffusive limit. Our simulation has not been long enough to produce good 
averaging in that domain. Note that this subdiffusive behavior is 
well known in supercooled liquids and also has been found in calamitic 
systems studied with the Gay-Berne pair potential~\cite{jose:chakrabarti:bagchi:pre:2005:030701}.

\subsection{Non-Gaussian parameter}

Non-Gaussian parameters (NGP) are often useful for description of the dynamical heterogeneity
in complex systems~\cite{schilling:rotation:supercool, leporini:supercool:pre}. 
For a system with linear rotors, the rotational NGP can be 
defined as~\cite{schilling:rotation:supercool, leporini:supercool:pre, 
jose:chakrabarti:bagchi:pre:2005:030701}:
\begin{equation}
\alpha_{2} = \frac{\langle \Delta \phi(t)^{4}\rangle}{2\langle \Delta \phi(t)^{2}\rangle^{2}} - 1,
\end{equation}
where
\begin{equation}
\langle \Delta \phi(t)^{2n}\rangle = \frac{1}{N} \sum_{i=1}^{N}\langle|\vec{\phi}_{n}(t) - \vec{\phi}_{n}(0)|^{2n}\rangle
\end{equation}

\begin{figure}
\centering
\includegraphics[width=0.47\textwidth]{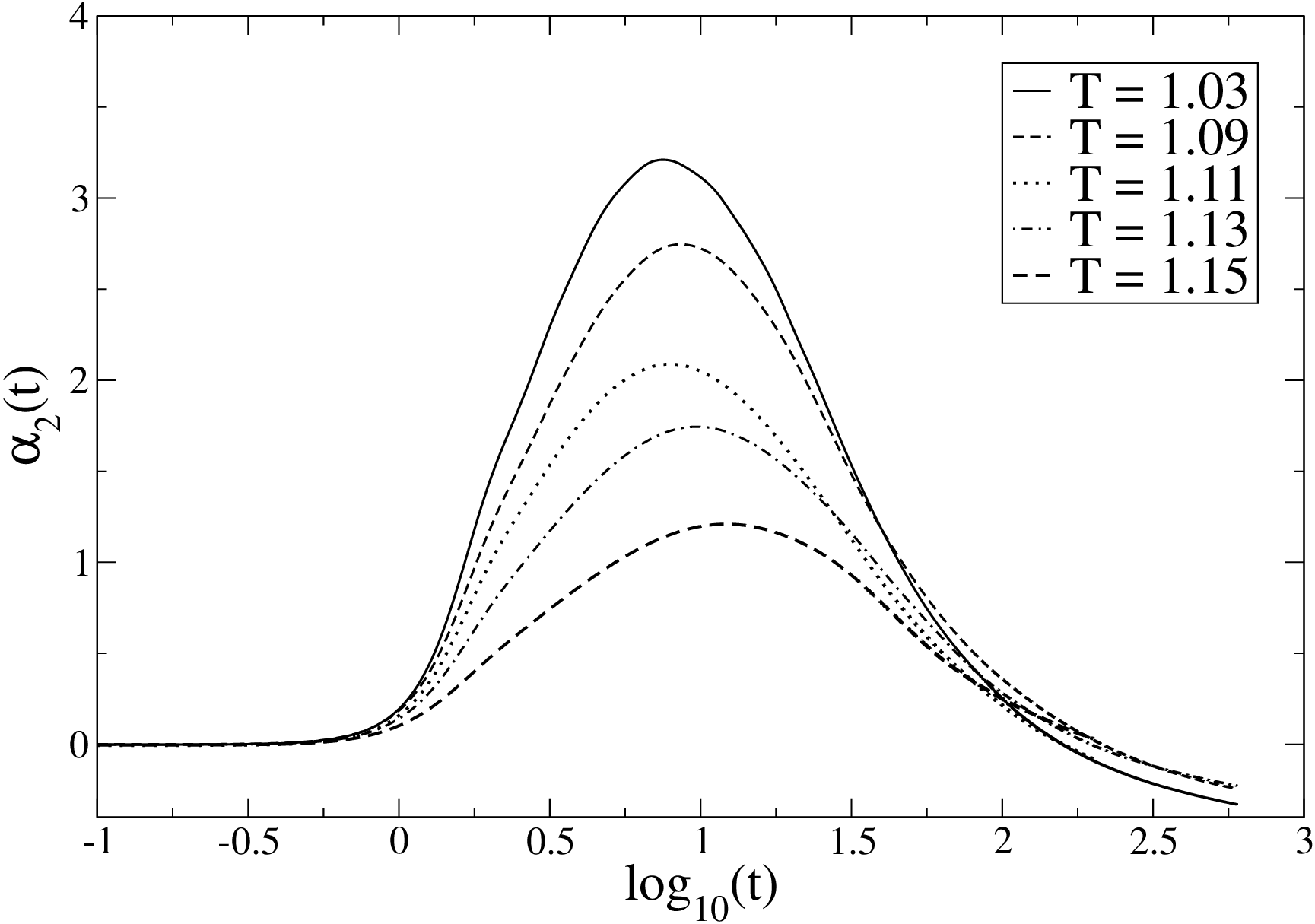}
\caption{\label{fig:ngp} The rotational non-Gaussian parameter across I-N transition shows 
the increasing non-Gaussian behavior in the nematic phase. We don't observe any appreciable 
regular shift of the maxima with temperature as in supercooled liquids.}
\end{figure}

In Fig.~\ref{fig:ngp} we show the time evolution of the rotational NGP. 
We observe large non-Gaussian behavior at intermediate timescales 
particularly at lower temperatures. Initially, when the motion of the rotors is ballistic, 
the rotational NGP is uniformly zero. It starts to grow at the comparable timescale as 
the appearance of the 
subdiffusive motion and the maximum is reached when the rotors escape from the orientational 
cage and gradually reaches the diffusive limit at longer timescales. The 
formation of pseudo-nematic domains as $T_{IN}$ is approached from above would make the 
system dynamically heterogeneous. The growing peak in $\alpha_{2}^{R}(t)$ can, therefore, 
be ascribed to the formation of pseudo-nematic domains. Note that similar behavior is well known 
in supercooled liquids~\cite{kob:andersen:pre, schilling:rotation:supercool, 
leporini:supercool:pre, jose:chakrabarti:bagchi:pre:2005:030701} and has been shown 
to be present in lattice models, e.g. in relaxation 
of phenomenological Brownian rotors based on the densely frustrated XY model~\cite{sqlattice:XYmodel}. 
However, we do not observe any regular shift in the position of the maximum within the 
temperature range of our consideration. 

\section{Free energy as a function of order parameter ($S$)}
\label{tmmc}

We have computed the free energy per rotor as a function of orientational order parameter using 
a variant of \emph{transition matrix Monte Carlo} (TMMC) method recently proposed by Fitzgerald 
\textit{et al}~\cite{fps:tmmc}. This method is considerably different than the 
reweighting methods for calculation of free energy and can be incorporated 
in any existing Monte Carlo simulation. An outline of the TMMC method applied to 
the present problem follows.

The basic idea is to calculate the probability ($\Pi(S,\beta)$) that the system is in a 
macrostate $S$ (in our case, average orientational order parameter) at the inverse 
temperature $\beta$. Note that any average 
observable of the system can be computed once the macrostate probability is known, 
e.g.\ in canonical ensemble, the free energy can be 
obtained as a function of order parameter from $\beta F(S,\beta) = - \ln \Pi(S,\beta)$. The 
\emph{Boltzmann macrostate probability} can be expressed as $\Pi(S,\beta) = \sum_{s \in S} \pi(s,\beta)$, 
where $\pi(s,\beta)$ is the Boltzmann probability for a particular microstate (microscopic 
configuration) $s$. It is given by $\pi(s,\beta)=\exp (-\beta H_{s})$, where $H_{s}$ is the 
value of the Hamiltonian for the microstate $s$.

The algorithm of finding $\Pi(S, \beta)$ is as follows:
\begin{enumerate}
  \item Similar to Metropolis algorithm, for a given initial microstate $s$, a new state 
$t$ is proposed with probability $q_{s,t}$. As a simplification, we chose $q_{s,t} = q_{t,s}$ 
though it is not strictly necessary.
  \item The probability of the move to state $t$ being accepted is 
\begin{equation}
r_{s,t}(\beta, \eta) = min \{1, \frac{\exp(\eta_{T}) \pi(t,\beta)}{\exp(\eta_{S}) \pi(s,\beta)} \},
\end{equation}
where $\eta_{S}$ is the weight function corresponding to macrostate $S$. For our 
purpose, we have set $\eta_{S} = -\ln \Pi(S, \beta)$, which 
corresponds to the \emph{multicanonical approach}~\cite{fps:tmmc}. Thus the macrostates with lower 
probability are given more weight so that toward the end of the simulation the 
histogram becomes flat and low probability states are sampled quite well.
  \item A new bookkeeping step is incorporated following the equilibration of the above 
Markov chain. At every step an array $C_{S,T}$ (initialized to zero) 
is incremented as follows:

  For $S \neq T$,
\begin{eqnarray}
  C_{S,T}(\beta) & = & C_{S,T}(\beta) + r_{s,t}(\beta, \eta = 0) \nonumber \\
  C_{S,S}(\beta) & = & C_{S,S}(\beta) + (1 - r_{s,t}(\beta, \eta = 0))
\end{eqnarray}

  For $S = T$, $C_{S,S} = C_{S,S} + 1$. Note that while the Markov chain is guided by 
the multicanonical weight, the unweighted Boltzmann transition probabilities are stored for 
each visited macrostate.
\end{enumerate}

The \emph{canonical transition probability} (CTP) between the macrostates has been calculated at 
some interval as
\begin{equation}
  P_{S,T}(\beta) = \frac{C_{S,T}(\beta)}{\sum_{U} C_{S,U}(\beta)}
\end{equation}

The equilibrated Markov chain must obey the detailed balance equation 
$\Pi_{S}(\beta) P_{S,T}(\beta) = \Pi_{T} P_{T,S}(\beta)$. Hence, the 
macrostate probabilities have been obtained by solving the set of coupled 
linear equations iteratively. The weights have been updated 
at every $150 \times L^{3}$ steps and the simulations have been continued for 
$5 \times 10^{6} \times L^{3}$ steps. The system learned to pass through the 
low probability states automatically and reasonably good sampling was attained.

\begin{figure}
\centering
\includegraphics[width=0.47\textwidth]{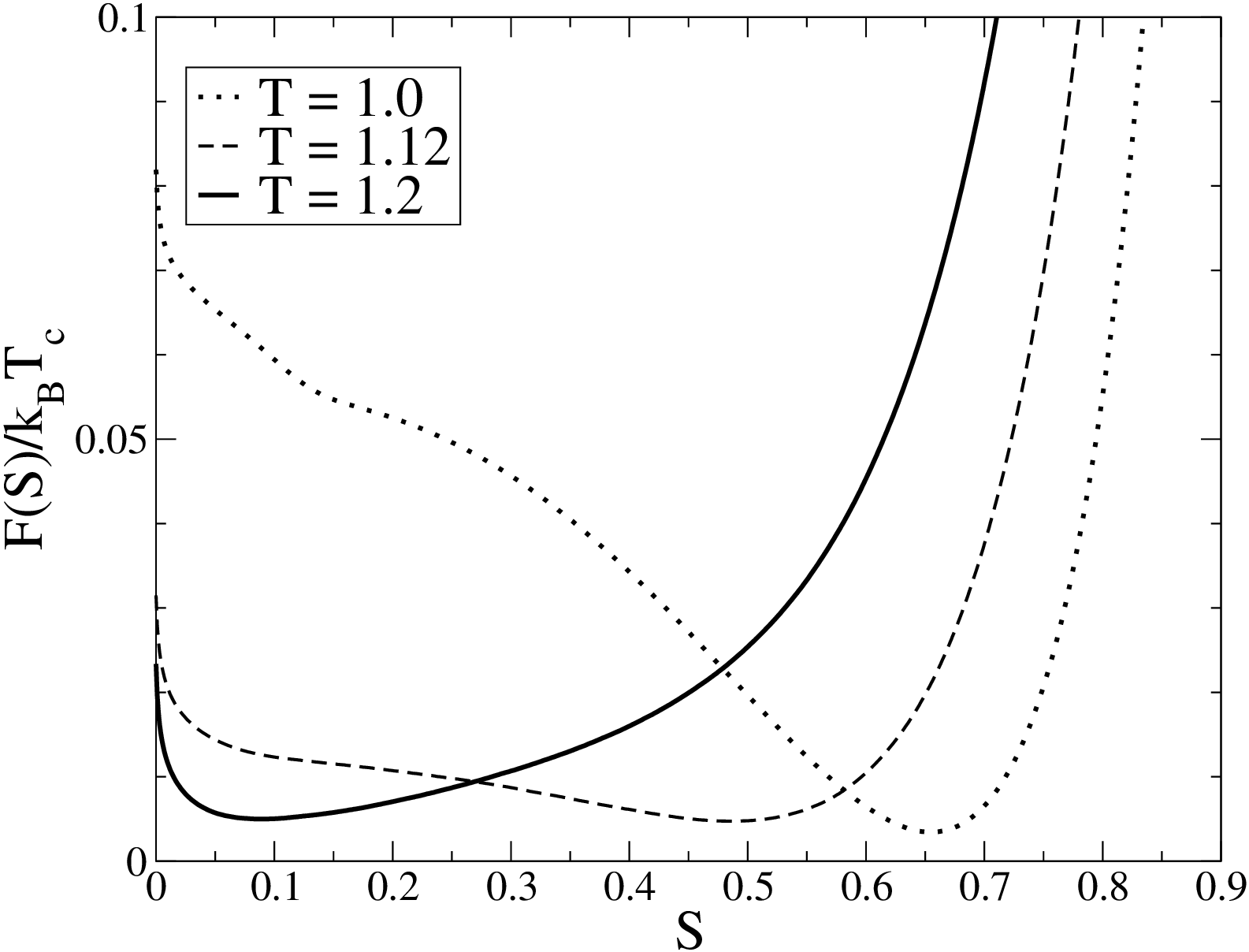}
\caption{\label{fig:freenrg} Free energy versus order parameter plot for 
a $10 \times 10 \times 10$ system at three different temperatures.}
\end{figure}

The computed free energy surface is shown in Fig.~\ref{fig:freenrg}. The flatness of the 
free energy surface near $T_{IN}$ is certainly a consequence of the \emph{very weakly
first order} nature of the transition and also partly due to the finite size of the
system~\cite{ll_freenrg:prl:1992}. Thus, a system of this size exhibits large scale 
fluctuation in the value of the orientational order parameter. Such fluctuations can 
give rise to the power law decay in OTCFs~\cite{liqcrys:prl:2005}. As the system 
size increases, a small barrier separating
the isotropic and nematic phases appears near the critical 
point~\cite{ll_freenrg:prl:1992}. 
This barrier height is directly related to the surface tension between the isotropic and 
the nematic phases and it follows a finite size scaling law~\cite{binder:surface, 
ll_freenrg:prl:1992, fps:tmmc}:
\begin{equation}
  \frac{F_{s}}{k_{B}T} = \lim_{L \rightarrow \infty} \frac{1}{2L^{d-1}}
  \ln \left[\frac{\Pi_{max}(\beta)}{\Pi_{min}(\beta)}\right],
\end{equation}
where $F_{s}$ is the surface free energy, $d$ is the dimension, $L$ is the box length, 
and $\Pi_{max}(\beta)$ and $\Pi_{min}(\beta)$ correspond to the maximum and minimum 
values of macrostate probability respectively.

Our result for 1000-particle system differs from one of the earlier 
reports~\cite{ll_freenrg:prl:1992}, which uses Ferrenberg-Swendsen reweighting technique,
as the positions of the minima are comparatively well-separated in our case while we consider 
temperatures far away from $T_{IN}$. We feel this has better agreement with the 
average order parameter values that one should obtain at those temperatures.
But for complete comparison between the methods and to verify the scaling relations 
for free energy barrier, more detailed study in larger systems is necessary.

It was discussed previously that the nearly flat free energy surface can be included 
in a Ginzburg-Landau type free energy functional~\cite{liqcrys:prl:2005}. One can 
then write down the following generalized Langevin equation for the fluctuating 
nonconserved order parameter $(\delta S)$~\cite{hohenberg:halperin, liqcrys:prl:2005}:
\begin{equation}
\label{genlangeq}
\frac{d (\delta S)}{dt} = - \int dt' \Gamma (t-t') \frac{\delta F}{\delta (\delta S)}(t') + R(t),
\end{equation}
where $\Gamma$ is a damping coefficient, $F(\delta S)$ is the Landau-de Gennes 
free energy as a function of the orientational order parameter and $R(t)$ is a 
random velocity term related to $\Gamma$ by the fluctuation-dissipation theorem. 
As the temperature approaches the critical temperature $T_{c}$, the free energy 
surface becomes soft. If one uses the Landau free energy expansion 
\begin{equation}
\label{freenrg:expansion}
  \delta F = A(T)(\delta S)^{2} + B(T) (\delta S)^{3} + C(T) (\delta S)^{4},
\end{equation}
then it can be shown that Eq.~\ref{genlangeq} can give rise to a power law 
decay of $\frac{<\Delta \bar{S}(t)>}{<\Delta \bar{S}(0)>}$ at short to intermediate times.

If the noise term in Eq.~(\ref{genlangeq}) is neglected and a Markovian 
approximation is made of $\Gamma (t)$, then one finds a power law 
in the decay of $\langle \delta S(0) \delta S(t) \rangle$. This power law 
originates from the cubic and the quartic terms in the free energy expansion 
given in Eq.~(\ref{freenrg:expansion}). However, an analytic solution in 
presence of the noise term $R(t)$ becomes highly non-trivial and needs to 
be carried out numerically. Work in this direction is in progress.

There are, however, two limits where one can obtain semi-quantitative answer directly. 
If the free energy surface is nearly flat, as shown in Fig.~\ref{fig:freenrg}, then decay 
of $\langle \delta S(0) \delta S(t) \rangle$ occurs via a generalized diffusion. In this case, 
the power law decay originates from power law behavior (in time or frequency) of 
$\Gamma (t)$. Note that flatness of free energy surface implies that $A(T) \approx 0$ near 
$T_{c}$. A study of relaxation by diffusion with a non-Markovian memory function was reported 
by Denny \textit{et al}~\cite{denny:bagchi:1999}. The alternate limit is where a 
significant barrier develops between the isotropic and the nematic phases. This, however, 
is expected to give largely an exponential decay. 

Note that the Landau-de Gennes behavior is obtained by keeping only the quadratic term in 
Eq.~(\ref{freenrg:expansion}) and the relaxation time is given by (in present notation):
\begin{equation}
\label{ldg:time}
  \tau_{LdG} = \frac{1}{2 \Gamma (z = 0) A(T)}
\end{equation}
The exponential relaxation observed both in experiments and simulations follow the 
Landau-de Gennes behavior with time constant given by Eq.~(\ref{ldg:time}).

\section{Conclusion}
\label{conclusion}

This report contains to the best of our knowledge the first detailed study of dynamics in LL 
model near the isotropic-nematic transition in the context of power law relaxation. 
Despite having a very simple intermolecular potential and no translational degrees of freedom,
the LL model exhibits an array of interesting dynamical features near the I-N 
phase boundary. The orientational relaxation slows down at intermediate timescales
possibly due to the \emph{caging effect} produced by neighboring sites and 
formation of pseudo-nematic domains in the isotropic phase very close to $T_{IN}$. The caging causes the
rotors to show subdiffusive behavior. But the caging is not strong enough and the system remains
underdamped over a long period of time. Essentially this leads to the vanishingly small free
energy barrier separating the isotropic and the nematic phases confirming the nature of the phase
transition to be very weakly first order.

It is interesting to note the synergy in the emergence of the power law decay in 
orientational relaxation and the
sub-diffusive behavior of the rotational diffusion. It is also equally
interesting to find the super-diffusive behavior that follows the sub-diffusive behavior.
\emph{This signifies the long flights of the rotors just after they escape from the cage.}

The analogy between the relaxation in supercooled liquids and that in the isotropic phase
near the I-N phase transition has been discussed only in recent literature~\cite{cang:fayer:jcp:2003, 
fayer:jcp:2006, jose:chakrabarti:bagchi:pre:2005:030701}. 
Even the present simple model shows dynamic signatures which are well-known in supercooled
liquids. Prominent among them, are the power law relaxation and the sub-diffusive behavior.
It is interesting to compare with the scenario in supercooled liquid where one often finds 
two power laws, one leading towards the plateau and the second one (Von Schweidler law) at 
longer times moving away from the plateau~\cite{cang:fayer:jcp:2003}. However, the origin 
of such analogous behavior in the two systems can be quite different. In the case of liquid crystals,
the onset of long range orientational correlations and the formation of pseudo-nematic domains
are responsible for the power law. Anomalies in the relaxation behavior in supercooled liquids, 
on the other hand, are believed to have kinetic origin.

There now appear to exist two somewhat different interpretations of the power law decay. The 
first one, proposed by Gottke \textit{et al}~\cite{fayer:bagchi:jcp:2002:360, 
fayer:bagchi:jcp:2002:6339}, is in terms of diverging orientational pair correlation function 
giving rise to a power law divergence of the memory function at low frequency 
(Eq.~(\ref{memory:powerlaw})). The alternative explanation offered recently by Bagchi and 
coworkers~\cite{liqcrys:prl:2005}, invokes large scale fluctuations in collective 
orientational density. The first explanation (in the spirit of MCT) does not require 
such large scale density fluctuation. It does, however, invoke diverging correlation length. 
The second explanation is particularly relevant for weakly first order transitions with 
second order characteristics where the two phases are separated by low barrier.

Because of the length of the MD simulations required to obtain a reliable, statistically
significant power law decay, we have been limited to study a 1000-particle system. The
underdamped nature of the relaxation has made the statistical averaging demanding. However, it
would be worthwhile to consider larger systems. We have made preliminary study of a 
$(20 \times 20 \times 20)$ lattice system (8000 particles). We find that the 
relaxation behavior does not change significantly, but the free energy surface starts 
showing the formation of a noticeable (but still small) barrier at the intermediate 
values of the order parameter as expected~\cite{ll_freenrg:prl:1992}.

\begin{acknowledgments}
It is a pleasure to thank Dr.\ P.\ P.\ Jose and Dr.\ P.\ Bhimalapuram for helpful 
discussions. This work was supported in part by grants from DST, India and CSIR, India. 
S.\ C.\ and D.\ C.\ acknowledge CSIR, India and UGC, India, respectively, for providing 
financial support.
\end{acknowledgments}

\bibliography{liqcrys}

\end{document}